# Media Usage in Post-Secondary Education and Implications for Teaching and Learning

G. Gidion[1], L.F. Capretz[2,*], K.N. Meadows[2] and M. Grosch[1]

[1]Karlsruhe Institute of Technology, Germany
[2]University of Western Ontario, Canada

## Abstract

The Web 2.0 has permeated academic life. The use of online information services in post-secondary education has led to dramatic changes in faculty teaching methods as well as in the learning and study behavior of students. At the same time, traditional information media, such as textbooks and printed handouts, still form the basic pillars of teaching and learning. This paper reports the results of a survey about media usage in teaching and learning conducted with Western University students and instructors, highlighting trends in the usage of new and traditional media in higher education by instructors and students. In addition, the survey comprises part of an international research program in which 20 universities from 10 countries are currently participating. Further, the study will hopefully become a part of the ongoing discussion of practices and policies that purport to advance the effective use of media in teaching and learning.

## 1. Introduction

Students tend to be early adopters of media and information technology, as they possess ample opportunities to access media, encouraged by their curiosity and self-taught skills. But students are not just passive users of technology; they are also designers and developers of technology. For example, Stanford students created Google, the most commonly used search engine on the Internet, in the latter Facebook, which was created by Harvard University students in 2004, has become one of the most successful Internet services worldwide in less than ten years.

The integration of IT media and services in higher education has led to substantial changes in the ways in which both students and instructors study, learn, and teach [1]. Accordingly, a survey of students' and instructors' media usage habits was conducted at Western University in 2013. This survey sought to measure the extent to which media services are used in teaching and learning as well as to assess changes in media usage patterns. The survey is a landmark, as it is the first of its kind in Canada and represents an initial foray into media usage habits of students and instructors in North America post-secondary sector. The study focuses on assessing the way in which media use relates to academic teaching and learning. . The identification of trends aims to provide an evidence base upon which future trends of media usage in higher education can be predicted more reliably. The basic hypothesis is twofold: Firstly, that current academic education is utilizing (and influenced by) a combination of traditional (e.g., printed books and journals) and new (e.g., Google and Wikipedia) media. Secondly, the current situation has developed from former media usage habits, and these habits might change with the introduction of new media. Future academic education will likely be influenced by media usage habits currently on the increase. The framework of this survey is depicted in Figure 1.

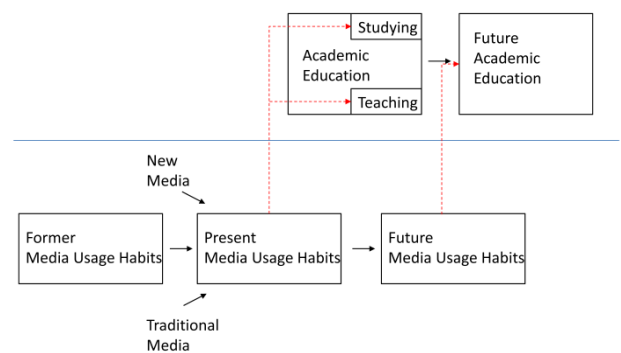

**Figure 1.** Framework of the media usage survey

*Corresponding author. Email: lcapretz@uwo.ca

In this study, media is defined as technology that supports and extends human communication. Information represents a unidirectional form of communication and, therefore, information services are also understood as media services. In the field of digital media, where the content lacks attachment to a physical data carrier, media services include software as well as hardware services. Because software media can be transferred to different hardware, the latter is necessary for software access and, thus, hardware constitutes an integral component of the definition of media. The Media Usage Survey was created to provide educational researchers with a deeper and more detailed understanding of students' and instructors' technology usage in learning contexts and of possible environmental factors that may influence that usage. This survey intended to incorporate the entire spectrum of media services, focusing on the following objectives:

- Evaluating media use in detail, including media use frequency and satisfaction with, and acceptance of, both internal or university-provided and external services, print media, electronic text, social media, information technology, communication media, e-learning services, and IT hardware;
- Determining factors that might influence media use in learning, such as cultural, age, sex, and academic level differences as well as identifying similarities among student media usage;
- Creating a knowledge base for universities to understand the media usage of students and instructors as well as establishing a longitudinal international survey on technology use in tertiary education;
- Assessing prospective media trends and supporting the definition of media development as one of the strategic ideas at universities;
- Evaluating user satisfaction, and thus media quality, by measuring the acceptance of services used by students and instructors.

## 2. Current media research

One of the most comprehensive media surveys to date was conducted by the EDUCAUSE Center for Applied Research (ECAR) in the Study of Undergraduate Students and Information Technology 2012 [2]. EDUCAUSE describes itself as "a nonprofit association and the foremost community of IT leaders and professionals committed to advancing higher education" and specifies that "EDUCAUSE programs and services are focused on analysis, advocacy, community building, professional development, and knowledge creation because IT plays a transformative role in higher education."

Some of the key highlighted findings indicate that blended learning environments seem to be more and more the norm, a change that most students welcome as the best support for their learning. Students expect their instructors to use technology to engage them in the learning process. For example, the study asked students about their interest in working with open educational resources. According to EDUCAUSE [2], in 2012, 57% of students said they wanted their instructors to use freely available course content more frequently, a number that had substantially increased from 19% the previous year. The emergence of freely available content is part of the way open solutions are transforming higher education. Examples of open educational resources include the Open CourseWare Consortium and the Khan Academy.

The Horizon Report of the New Media Consortium [1], which is related to the EDUCAUSE study, concentrates on future trends. In the 2013 report, they differentiate between perspectives for one year: MOOCs and tablet computing; two to three years: learning analytics and games and gamification; and four to five years: wearable technology and 3D printing.

For the current year, the report highlights the introduction of tablet computing and Massive Open Online Courses (MOOC). MOOCs received their fair share of hype in 2012 and are expected to grow in number and influence over the next year. Big name providers, including Coursera, edX, and Udacity, each count hundreds of thousands of enrolled students, totals that illustrate their popularity when combined. One of the most appealing promises of MOOCs is that they offer the possibility for continued, advanced learning at no cost, allowing students, life-long learners, and professionals to acquire new skills and improve their knowledge and employability. MOOCs have enjoyed one of the fastest uptakes ever seen in higher education, with literally hundreds of new courses added in the last year. However, critics loudly warn that there is a need to examine these new approaches through a critical lens to ensure they are effective and evolve past the traditional lecture-style pedagogies. In the near future, the report expects games and gamification and learning analytics to increase in popularity and use; the more distant future is expected to be most influenced by wearable technology and 3D printing.

According to Buckingham [3], students' everyday use of computer games, mobile devices, and the Internet involves a range of informal learning processes, in which participants are simultaneously teachers and learners. Participants learn to use these media largely through trial and error, that is, exploration, experimentation and play, and collaboration with others in both face-to-face and virtual forms. Buckingham [3] asserts that one cannot teach about contemporary media without taking into account the role of the Internet, computer games, and the convergence between old and new media. Much of the popular discussion in this area tends to assume that contemporary students already know everything about new media; they are celebrated as "millennials," or as "digital natives" who are somehow spontaneously competent and empowered in their dealings with new media.

Traditional forms of teaching, which involve the transmission of a fixed body of information, are largely irrelevant nowadays. This is particularly evident with the more recent rise of participatory media in the form of social networking sites, file sharing, wikis, and blogs. Other technology-enhanced lectures have been put forward by several researchers on this topic, such as LearnWeb2.0 [4] and LaaN [5]. However, considering the popularity and ubiquitous nature of these new technologies, particularly the potential of mobile learning, their acceptance in educational institutions is still considered limited [6].

## 3. Motivation

Students in post-secondary education intensively use web services, such as Google, Wikipedia, and Facebook during their free time as well as for their studies [7]. Current development in the so-called web 2.0 is often characterized by an increase in interactions between users, as seen in the rise of collaborative and participatory media in the form of social network software.

Pritchett et al. [8] examined "the degree of perceived importance of interactive technology applications among various groups of certified educators" (p. 34) and found that, in the involved schools, participants of the survey with "an advanced degree and/or higher certification level" (p. 37) seemed to perceive Web 2.0 media as more important than other participants did. Furthermore, mobile broadband Internet access and the use of corresponding devices, such as netbooks and smartphones, have fuelled the use of social networks by students in higher education. Murphy et al. [9] reported that in spite of the limitations in formal university infrastructure, many students would like to use their mobile devices for formal as well as informal learning. Relatedly, recent developments in technology resulting in smartphones and tablets dominating the market in recent years have ensured that these devices have great functionality and enable interactivity, thus fulfilling the desire for both formal learning [10] and informal learning [11].

There have been doubts about the potential of this technological shift in students' learning and the real benefits of these technologies for learning. Considerable research has outlined the costs and benefits of using social, mobile, and digital technology to enhance teaching and learning; yet the research is not conclusive as to whether the use of these technologies leads to improved learning outcomes [12]. Klassen [13] stated. "If there is one thing I have learned the last ten years about the use of new technology in education, it is that the combination of old and new methods make for the best model" and went on to say "Students will continue to seek out inspiring teachers. Technology alone is unlikely to ensure this, although it may make a lot of average teachers seem a lot better than they are!"

The usage of media at university is a topic of interest for students, staff, and faculty. While there may be diverse interests and habits, several interdependencies and interactions exist. The understanding of one of these scenarios was the objective of a study by Kazley et al. [14], who surveyed students, staff, and faculty and defined certain "factors that determine the level of educational technology use" (p. 68). They described a model with increasing intensity and quality of technology use, from beginners (using e-mail and basic office software) to experts (using videoconferencing, virtual simulation tools, etc.).

There is no doubt, however, that the integration of IT media and services in higher education appears to have led to substantial changes in the ways in which students study and learn. Higher education institutions are cautious about investing in programs to provide students with mobile devices for learning due to the rapidly changing nature of technologies [15]. The acceptance of technology-enhanced education by students has increased in recent years, but not all services are equally accepted [16]. It has become clear that simply using media and adopting e-learning does not necessarily make a difference in student learning. Rather, pedagogy and the quality of the services are key factors for the effective use of technology [17], [18], [19].

Also, the variety of media enriched informal learning processes is relevant. This perspective on the whole spectrum of media used for learning (printed, e-learning, digital, web 2.0, etc.) requires a certain theory-oriented empirical research approach to reach a deeper understanding about the media usage behavior of higher education students.

## 4. Research methodology

The survey comprised a fully standardized anonymous questionnaire containing a total of 150 items. Specifically, the tool measured usage frequency and user satisfaction with 53 media services, including:

- Media hardware and web connection, such as Wi-Fi, notebooks, tablet computers, desktop computers, and smartphones;
- Information services, such as Google search, Google Books, library catalogues, printed books, e-books, printed journals, e-journals, Wikipedia, open educational resources, and bibliographic software;
- Communication services, such as internal and external e-mail, Twitter, and Facebook;
- E-learning services and applications, such as learning platforms and wikis.

These variables, as well as the previously mentioned methodology, were also used to create acceptance values. Additional variables underwent evaluation, such as some aspects of learning behavior, media usage in leisure time, educational biography, and socio-demographic factors. There were several groups of questions, as represented in Figure 2.

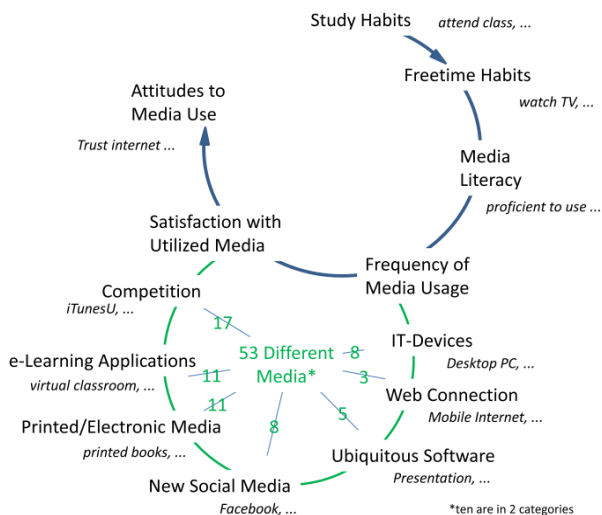

**Figure 2.** Schematic image of the survey's main groups of questions

The **usage of diverse IT devices** is represented with the items: desktop PC, computer labs on campus (e.g., Genlab), own notebook/laptop off campus, own notebook/laptop on campus, mobile phone (e.g., smartphone, iPhone, Blackberry, Samsung), tablet computer (e.g., iPad, Galaxy Tab, Blackberry PlayBook), E-book reader (e.g., Kindle, NOOK, Sony Reader), and gesture computing (Xbox Kinect, IPhone interface, Nintendo Wii).

The **usage of variable web connections** is represented with the items: mobile Internet connection (with notebook, tablet, phone), Internet connection at home and wireless connection (Wi-Fi, WLAN) on campus.

The **usage of various software** is represented with the items: learning software, educational software, dictionary software installed on your computer, bibliographic software (e.g., Endnote, Zotero, RefWorks), word processing software, and presentation software (PowerPoint, Keynote, Prezi).

The **usage of e-Learning applications** is represented with the items: e-learning applications as part of a course, wikis with active participation as part of a course, online materials from other universities (e.g., iTunesU, Coursera, MIT Open Courseware), learning software, educational software, recorded lectures (audio, video), online exams (for grades in a course), online (self) tests for studying, video sharing websites (e.g., YouTube), game-based learning applications, augmented reality application (e.g., Geotagging in Google Earth), and mobile apps for learning (e.g., iTunesU, iBooks).

The **usage of social network related applications** is represented with the items: newsgroups, Internet forums, Wikipedia, Google search, social bookmarking and tagging (e.g., Delicious), Facebook, Google+, other social networks sites (e.g., LinkedIn), and Twitter.

The **usage of university-intern media vs. media offered by external providers** is represented with the items: online materials from other universities (e.g., iTunesU, Coursera, MIT open-courseware), university websites (e.g., the website of the University of Western), web portal for online student web services (e.g., PeopleSoft), learning management system (e.g., Sakai/OWL, Moodle), online slides (e.g., PowerPoint, Keynote, Prezi) from an instructor, online material (lecture notes) and/or scientific articles from an instructor, recorded lectures (audio, video), virtual class in real time (virtual lectures, web conferences), virtual class in non-real time/asynchronous (web seminars, webinars), printed handouts from an instructor, online services at the university library (central)/faculty library, online services at other libraries (not university university), printed books, university e-mail account, e-mail account not associated with the university (e.g., Hotmail, Yahoo, Gmail), Google+, and instant messaging (e.g., MSN/Yahoo Messenger, Skype).

The **usage of printed vs. electronic digital media** is represented with the items: online dictionary, online slides (e.g., PowerPoint, Keynote, Prezi) from an instructor, online material (lecture notes) and/or scientific articles from an instructor, printed handouts from an instructor, printed books, ebooks (e.g., pdf, ePub, Mobi, Kindle, Kobo), print-versions of academic periodicals/journals, e-versions of academic periodicals/journals, Wikipedia, Google Books, and Google+.

The survey tool was first developed in 2009 and used at the Karlsruhe Institute of Technology (KIT) in Germany [20], [21]. During the application of the 15 follow-up surveys that were administered internationally, the original survey underwent optimization, translation into several languages, and validation. In this study, the survey was administered at Western University to undergraduate students and faculty members in January and February of 2013. The instructor survey and the student questionnaire intended to compare the media usage of students and instructors by examining possible divergences in media culture that may create problems in the use of media for studying and teaching.

Initial invitations to participate in the research and two reminders were sent by e-mail. Both faculty and student surveys were voluntary and anonymous, as indicated in the cover letters. For the student survey, three e-mails were sent by the Office of the Registrar staff to a stratified random sample of undergraduate and graduate students enrolled on the main campus in the Winter 2013 academic term. The faculty survey used a similar procedure and targeted faculty teaching on the main campus during the Winter 2013 academic term. The data for this survey was collected online using Unipark, an established survey provider. In the period between January 16 and February 15, 2013, 19,978 students were invited to respond to the survey. Subsequently, 1,584 visits occurred to the survey website. Among the invited students, 1,266 started to answer the questions, 985 completed the survey, and 803 recorded a completion rate of more than 90%. In the period between January 29 and February 28, 2013, approximately 1,400 instructors were solicited by e-mail to answer the survey. During this time, exactly 332 visits occurred to the survey website. Although 252 faculty members started to answer the questions, 210 of them completed the survey.

**Table 1.** Response numbers for Western students and instructors who indicated the faculty of their primary area of study or primary teaching assignment.

| | Students UG | | | | Instructors | | | |
|---|---|---|---|---|---|---|---|---|
| | Population | | Participants | | Population | | Participants | |
| | N | % | n | % (of 792) | N | % | n | % (of 187) |
| Arts and Humanities | 1,232 | 5.7 | 82 | 10.4 | 151 | 11.0 | 15 | 8.0 |
| Education | - | - | - | - | 37 | 2.7 | 4 | 2.1 |
| Engineering | 1,310 | 6.0 | 56 | 7.1 | 94 | 6.8 | 11 | 5.9 |
| Health Sciences | 3,246 | 15.0 | 125 | 15.8 | 133 | 9.7 | 21 | 11.2 |
| Information and Media Studies | 969 | 4.5 | 45 | 5.7 | 44 | 3.2 | 3 | 1.6 |
| Law | - | - | - | - | 33 | 2.4 | 2 | 1.1 |
| Music | 527 | 2.4 | 37 | 4.7 | 44 | 3.2 | 15 | 8.0 |
| Richard Ivey School of Business | 1,097 | 5.1 | 15 | 1.9 | 111 | 8.1 | 11 | 5.9 |
| School of Graduate & Postdoctoral Studies | - | - | - | - | - | - | 2 | 1.1 |
| Schulich School of Medicine & Dentistry | 2,425 | 11.2 | 19 | 2.4 | 281 | 20.5 | 42 | 22.5 |
| Science | 4,244 | 19.6 | 173 | 21.8 | 203 | 14.8 | 23 | 12.3 |
| Social Science | 6,627 | 30.6 | 237 | 29.9 | 241 | 17.6 | 38 | 20.3 |
| Missing (this item) | | | (193) | | | | (23) | |
| Total | 21,677 | | **985** | | 1372 | | **210** | |

While participants were randomly selected from a broad spectrum of demographic characteristics and faculties, female students were more heavily represented in terms of respondents [22]. Otherwise, with some caveats, respondents are generally regarded as representative of the January and February 2013 student and instructor population at Western. A summary of participation is shown in Table 1.

## 5. Main findings of the media usage surveys

Due to page limitations, partial results are presented in the subsequent self-explanatory figures and tables. In the study, usage frequency was connected with satisfaction with the media. The students who stated they used a media at any level of frequency were asked how satisfied they were with this usage. The questions were rated on a five-point Likert scale with the following choices: never (0), rarely (1), sometimes (2), often (3), and very often (4); very unsatisfied (0) to very satisfied (4).

The students were asked about their general media, learning, and studying habits. The results show that students most often attended class, followed by studying using a computer and studying by themselves at home. Searching on the Internet for learning materials seems to be slightly more common than visiting libraries. Compared to the other habits, cooperative learning seems relatively rare.

The items "attend class," "study by yourself at home" and "visit libraries" can be interpreted as indicators for activities that have been used since the foundation of universities. The items "study using a computer" and "search the Internet for learning materials" integrate relatively new activities into this group of "traditional" studying habits. The frequency of the latter items can be compared to that of the item "visit libraries" (which is probably used as an additional, not a substitute, activity) and "study with printed materials you found yourself" (which differentiates from material given by instructors). The items "study together with one other person" and "study in groups (more than two people)" are related to the item "study with other students online (via Facebook, Instant Messenger, or e-mail)" which exemplifies a new media-based option to cooperation and seems to be less common than conventional forms of joint studying. All three variations of joint studying were rated as less frequently realized than "isolated" learning arrangements. The results of all the eight items together generate the impression of a mixture between traditional and new general media, learning, and studying habits.

**Table 2.** Students' answers to the question: How often do you do the following?

| How often do you do the following? | Mean | Std. Dev. | valid n | Valid Percent | | | | |
|---|---|---|---|---|---|---|---|---|
| | | | | (0) never | 1 (rarely) | 2 (some-times) | 3 (often) | (4) very often |
| Attend class | 3.77 | 0.59 | 979 | 0.4 | 0.8 | 3.5 | 12.1 | 83.3 |
| Study by yourself at home | 3.21 | 1.00 | 974 | 0.8 | 7.2 | 15.0 | 24.2 | 52.8 |
| Search the Internet for learning materials | 2.98 | 1.01 | 971 | 0.9 | 9.1 | 19.3 | 32.8 | 38.0 |
| Study with printed materials you found yourself | 1.81 | 1.25 | 977 | 15.6 | 29.5 | 25.8 | 16.5 | 12.7 |
| Study together with one other person | 1.92 | 1.17 | 981 | 11.4 | 28.9 | 25.8 | 24.5 | 9.5 |
| Study in groups (more than two people) | 1.39 | 1.16 | 977 | 25.7 | 34.5 | 20.6 | 13.8 | 5.4 |
| Study with other students online (via Facebook, Instant Messenger, or e-mail) | 0.84 | 1.04 | 982 | 50.4 | 26.5 | 14.5 | 6.2 | 2.4 |
| Study using a computer | 3.26 | 1.02 | 980 | 2.2 | 5.1 | 13.1 | 23.3 | 56.3 |
| Visit libraries | 2.56 | 1.21 | 982 | 5.0 | 17.8 | 21.2 | 27.7 | 28.3 |

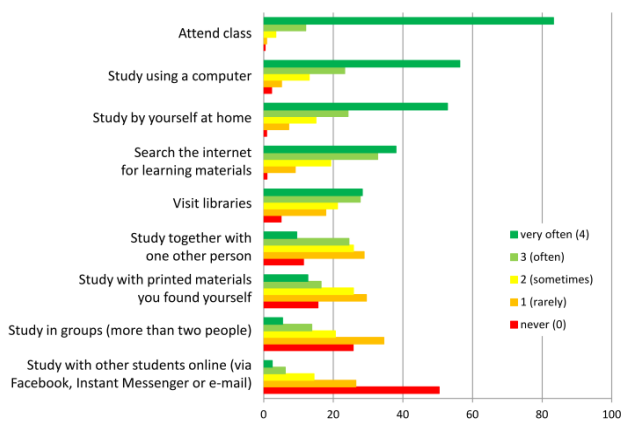

**Figure 3:** The valid percent of students' responses to the question: How often do you do the following? (ranking of items sorted by means)

The same group of items was used with minor modifications in the survey for instructors, and the results in this group show a slightly different picture. Instructors used the computer even more than students and Internet searches for teaching and learning materials were their second most frequent activity from this list of items. Cooperative work does not seem to happen as frequently as working alone, although it occurred more often for instructors than for students. This corresponds to a result in another group of items showing a higher frequency of usage of cooperative software by instructors than students.

**Table 3.** Instructors' responses to the question: How often do you do the following?

| How often do you do the following? | Mean | Std. Dev. | valid n | Valid Percent | | | | |
|---|---|---|---|---|---|---|---|---|
| | | | | (0) never | 1 (rarely) | 2 (some-times) | 3 (often) | (4) very often |
| Teach class | 3.29 | 1.020 | 207 | 1.0 | 7.7 | 12.6 | 19.3 | 59.4 |
| Work by yourself at home | 3.15 | 1.106 | 205 | 1.0 | 10.7 | 16.6 | 16.1 | 55.6 |
| Search the Internet for teaching or learning material | 3.36 | .886 | 210 | 1.0 | 3.8 | 10.5 | 28.1 | 56.7 |
| Learn with printed material you found yourself | 3.12 | 1.019 | 209 | 1.4 | 7.2 | 16.3 | 28.2 | 46.9 |
| Work together with one other person | 2.31 | 1.254 | 210 | 6.2 | 26.2 | 21.0 | 23.8 | 22.9 |
| Work in groups (more than two people) | 2.00 | 1.332 | 205 | 12.2 | 31.2 | 21.5 | 14.6 | 20.5 |
| Work with other colleagues online (via Facebook, Instant Messenger, or e-mail) | 1.99 | 1.388 | 208 | 17.3 | 24.5 | 20.7 | 17.3 | 20.2 |
| Work using a computer | 3.91 | .452 | 209 | .5 | .5 | 1.9 | 1.4 | 95.7 |
| Visit libraries | 2.03 | 1.240 | 209 | 8.1 | 33.5 | 23.0 | 18.2 | 17.2 |

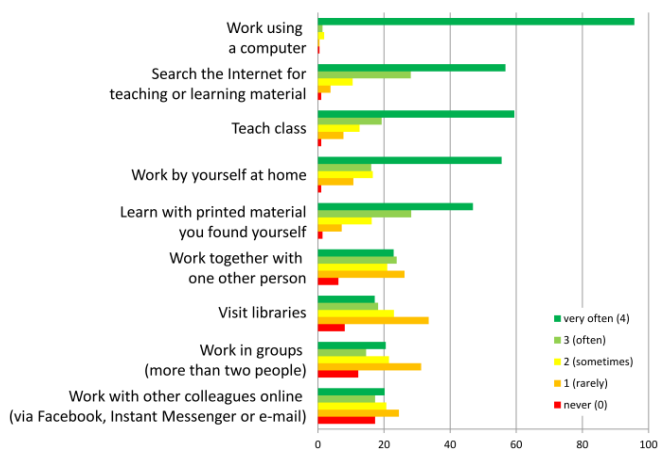

**Figure 4.** Valid percentage of instructors' responses to the question: How often do you do the following? (list of items sorted by means)

## Frequency of Diverse IT Devices Usage

Students were asked how often they use various IT devices for learning and studying. Most intensive use seems to be their own notebook or laptop off campus, although the intensity of this use was close to that of the use of the same equipment on campus. The use of mobile phones, such as a smartphone, iPhone, Blackberry, or Samsung, was less intensive, but it was still higher than the use of desktop PCs. Computer labs on campus were more in use than desktop PCs. Students from specific faculties, such as engineering, use these labs more than students from other faculties. Tablet computers or IT equipment that supports gesture computing seem to be in use less often. E-book readers were used less frequently by students.

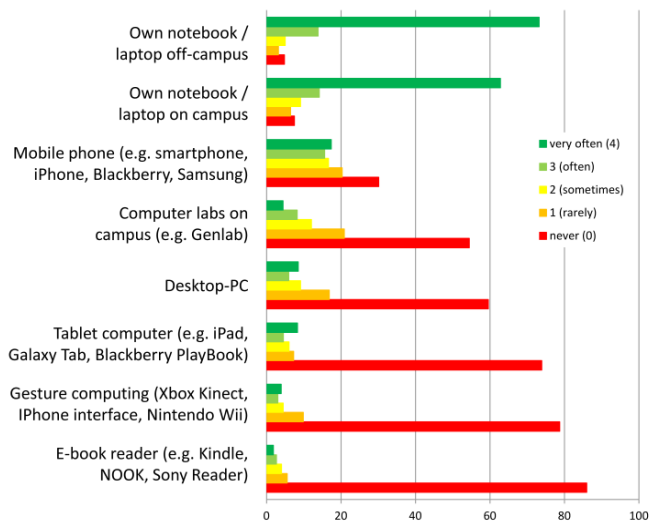

**Figure 5.** Valid percentage of students' responses to the question: How often do you use the following for learning/studying?

**Table 4.** Students' answers to the question: How often do you use the following for learning/studying?

| | How often do you use the following for learning/studying? | Mean | Std. Dev. | valid n | (0) never | 1 (rarely) | 2 (some-times) | 3 (often) | (4) very often |
|---|---|---|---|---|---|---|---|---|---|
| | | | | | colspan Valid Percent | | | | |
| 1 | Mobile phone (e.g., smartphone, iPhone, Blackberry, Samsung) | 1.70 | 1.47 | 980 | 30.1 | 20.3 | 16.6 | 15.6 | 17.4 |
| 3 | Own notebook/laptop off campus | 3.47 | 1.06 | 978 | 4.8 | 3.2 | 5.0 | 13.8 | 73.2 |
| 4 | Own notebook/laptop on campus | 3.18 | 1.27 | 976 | 7.5 | 6.5 | 9.1 | 14.1 | 62.8 |
| 5 | Desktop PC | 0.87 | 1.29 | 974 | 59.6 | 16.8 | 9.1 | 6.0 | 8.5 |
| 7 | Tablet computer (e.g., iPad, Galaxy Tab, Blackberry PlayBook) | 0.66 | 1.27 | 977 | 73.9 | 7.3 | 6.0 | 4.5 | 8.3 |
| 8 | E-book reader (e.g., Kindle, NOOK, Sony Reader) | 0.29 | 0.82 | 977 | 86.0 | 5.5 | 4.0 | 2.7 | 1.8 |
| 18 | Computer labs on campus (e.g., Genlab) | 0.87 | 1.17 | 972 | 54.4 | 20.9 | 12.0 | 8.2 | 4.4 |
| 52 | Gesture computing (Xbox Kinect, iPhone interface, Nintendo Wii) | 0.44 | 1 | 963 | 78.7 | 9.9 | 4.5 | 3.0 | 4.0 |

## Satisfaction with the Usage of Diverse IT Devices

In the survey, usage was connected with satisfaction related to the specific use of media. The students who stated they used a media frequently were asked how satisfied they were with this usage. The question was rated on a five-point Likert scale with the choices: never (0), rarely (1), sometimes (2), often (3), and very often (4); resp. very unsatisfied (0) to very satisfied (4).

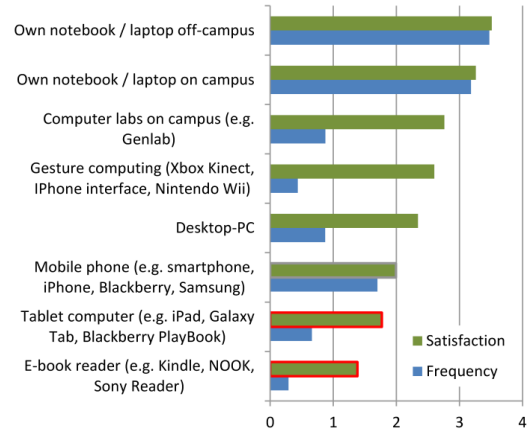

**Figure 6.** Means of students' responses to the questions: How often do you use the following for learning/studying? and If you use it: How satisfied are you with the use/functionality of the following for learning/studying? (red outline: in direction of dissatisfaction).

**Table 5.** Means of students' answers to the questions: How often do you use the following for learning/studying? and If you use it: How satisfied are you with the use/functionality of the following for learning/studying?

| | How often do you use the following for learning/studying? and If you use it: How satisfied are you with the use/functionality of the following for learning/studying? | Frequency | Satisfaction |
|---|---|---|---|
| 3 | Own notebook/laptop off campus | 3.47 | 3.51 |
| 4 | Own notebook/laptop on campus | 3.18 | 3.26 |
| 18 | Computer labs on campus (e.g., Genlab) | .87 | 2.76 |
| 52 | Gesture computing (Xbox Kinect, IPhone interface, Nintendo Wii) | .44 | 2.60 |
| 5 | Desktop PC | .87 | 2.34 |
| 1 | Mobile phone (e.g., smartphone, iPhone, Blackberry, Samsung) | 1.70 | 1.99 |
| 7 | Tablet computer (e.g., iPad, Galaxy Tab, Blackberry PlayBook) | .66 | 1.77 |
| 8 | E-book reader (e.g., Kindle, NOOK, Sony Reader) | .29 | 1.38 |

The comparison between the means of frequency and the means of satisfaction with the IT devices shows high values for the usage of one's own notebook/laptop both on and off campus; a low (and, for certain groups, sometimes high) usage of computer labs on campus with high satisfaction, where they were in use; a rare use of gesture computing devices, but high satisfaction in cases of use. The satisfaction value of mobile phones, which were utilized relatively often, was on a lower-middle level; the means of satisfaction for the usage of tablet computers and e-book readers tended towards dissatisfaction.

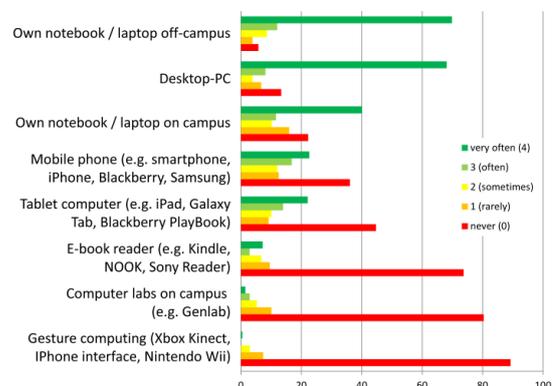

**Figure 7.** Valid percentage of instructors' responses to the question: How often do you use the following for your academic work (i.e., teaching, research, service)?

**Table 6.** Instructors' responses to the question: How often do you use the following for your academic work (i.e., teaching, research, service)?

| How often do you use the following for your academic work (i.e. teaching, research, service)? | | Mean | Std. Dev. | valid n | Valid Percent | | | | |
|---|---|---|---|---|---|---|---|---|---|
| | | | | | (0) never | 1 (rarely) | 2 (sometimes) | 3 (often) | (4) very often |
| 1 | Mobile phone (e.g., smartphone, iPhone, Blackberry, Samsung) | 1.77 | 1.613 | 208 | 36.1 | 12.5 | 12.0 | 16.8 | 22.6 |
| 3 | Own notebook/laptop off campus | 3.36 | 1.153 | 209 | 5.7 | 3.8 | 8.6 | 12.0 | 69.9 |
| 4 | Own notebook/laptop on campus | 2.31 | 1.638 | 207 | 22.2 | 15.9 | 10.1 | 11.6 | 40.1 |
| 5 | Desktop PC | 3.11 | 1.478 | 210 | 13.3 | 6.7 | 3.8 | 8.1 | 68.1 |
| 7 | Tablet computer (e.g., iPad, Galaxy Tab, Blackberry PlayBook) | 1.60 | 1.660 | 208 | 44.7 | 9.1 | 10.1 | 13.9 | 22.1 |
| 8 | E-book reader (e.g., Kindle, NOOK, Sony Reader) | .60 | 1.189 | 209 | 73.7 | 9.6 | 6.7 | 2.9 | 7.2 |
| 18 | Computer labs on campus (e.g., Genlab) | .35 | .825 | 209 | 80.4 | 10.0 | 5.3 | 2.9 | 1.4 |
| 52 | Gesture computing (Xbox Kinect, iPhone interface, Nintendo Wii) | .15 | .498 | 204 | 89.2 | 7.4 | 2.9 | .0 | .5 |

The results of the instructor survey show a higher value for the usage frequency of desktop PCs than that of students; it can be assumed that these devices were located in the instructors' offices. Fewer instructors worked with tablet computers, e-book readers, and gesture computing devices, and only few f instructors utilised computer labs on campus. In general, both students and instructors utilized mobile devices regularly.

## Frequency of e-Learning Applications Usage

Video sharing websites, such as YouTube, were only moderately used for learning purposes. Recorded lectures, audio and video and online self-tests for studying were both used rarely to moderately. Course-based e-learning applications and course-based wikis were rarely used, and mobile apps for learning, such as iTunesU and iBooks, and game-based learning applications were rarely to never used for learning. [Figure 8]

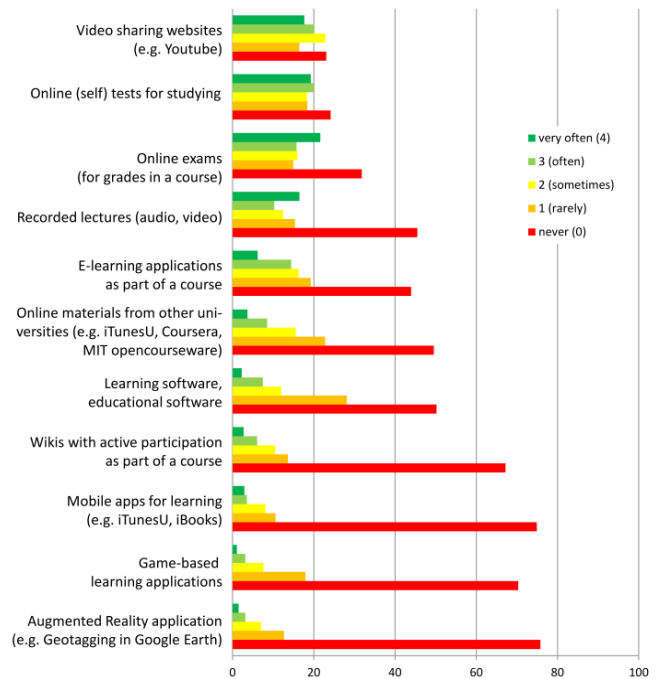

**Figure 8.** Valid percentage of students' responses to the question: How often do you use the following for learning/studying?

**Table 7.** Students' answers to the question: How often do you use the following for learning/studying?

| | How often do you use the following for learning/studying? | Mean | Std. Dev. | valid n | Valid Percent | | | | |
|---|---|---|---|---|---|---|---|---|---|
| | | | | | (0) never | 1 (rarely) | 2 (some-times) | 3 (often) | (4) very often |
| 9 | E-learning applications as part of a course | 1.20 | 1.3 | 972 | 43.9 | 19.2 | 16.3 | 14.4 | 6.2 |
| 11 | Wikis with active participation as part of a course | 0.63 | 1.06 | 963 | 67.2 | 13.6 | 10.5 | 6.0 | 2.7 |
| 12 | Online materials from other universities (e.g., iTunesU, Coursera, MIT OpenCourseWare,) | 0.94 | 1.15 | 979 | 49.5 | 22.8 | 15.5 | 8.5 | 3.7 |
| 13 | Learning software, educational software | 0.83 | 1.05 | 978 | 50.2 | 28.1 | 12.0 | 7.5 | 2.3 |
| 23 | Recorded lectures (audio, video) | 1.37 | 1.53 | 965 | 45.5 | 15.3 | 12.4 | 10.3 | 16.5 |
| 36 | Online exams (for grades in a course) | 1.80 | 1.55 | 959 | 31.8 | 14.9 | 16.0 | 15.8 | 21.6 |
| 37 | Online (self) tests for studying | 1.92 | 1.45 | 962 | 24.1 | 18.4 | 18.3 | 20.0 | 19.2 |
| 47 | Video sharing websites (e.g., YouTube) | 1.93 | 1.41 | 968 | 23.0 | 16.4 | 22.8 | 20.0 | 17.7 |
| 50 | Game-based learning applications | 0.47 | 0.85 | 960 | 70.3 | 17.9 | 7.6 | 3.1 | 1.0 |
| 51 | Augmented reality application (e.g., Geotagging in Google Earth) | 0.42 | 0.86 | 958 | 75.8 | 12.6 | 7.0 | 3.1 | 1.5 |
| 53 | Mobile apps for learning (e.g., iTunesU, iBooks) | 0.49 | 0.99 | 964 | 74.9 | 10.6 | 8.1 | 3.5 | 2.9 |

## Satisfaction with the Usage of e-Learning Applications

The results show high values of satisfaction even with the very frequently used items of online (self) tests for studying, online exams (for grades in a course), video sharing websites, and recorded lectures. In the middle field of this group of items are rarely utilized game-based learning applications and augmented reality applications, as well as the slightly more frequently used online materials from other universities and learning software. Wikipedia (an item in another group of questions in the survey) was used quite often, but work with wikis as an active participation method that is part of a course seems not only to be rarely utilized, but also not very satisfying from the perspective of the students involved in the study. A mean with a tendency to dissatisfaction (in a state of a not so low value of usage frequency) was the result concerning e-learning applications as part of a course and mobile apps for learning. Based on this, it appears that these applications might as well not be further developed and established for the time being.

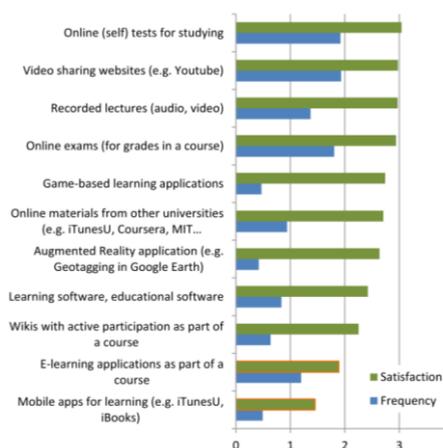

**Figure 9.** Means of students' responses to the questions: How often do you use the following for learning/studying? and If you use it: How satisfied are you with the use/functionality of the following for learning/studying? (red outline: in direction of dissatisfaction).

**Table 8.** Means of students' answers to the questions: How often do you use the following for learning/studying? and If you use it: How satisfied are you with the use/functionality of the following for learning/studying?

| | How often do you use the following for learning/studying? and If you use it: How satisfied are you with the use/functionality of the following for learning/studying? | Frequency | Satisfaction |
|---|---|---|---|
| 37 | Online (self) tests for studying | 1.92 | 3.04 |
| 47 | Video sharing websites (e.g., YouTube) | 1.93 | 2.97 |
| 23 | Recorded lectures (audio, video) | 1.37 | 2.97 |
| 36 | Online exams (for grades in a course) | 1.80 | 2.94 |
| 50 | Game-based learning applications | .47 | 2.74 |
| 12 | Online materials from other universities (e.g., iTunesU, Coursera, MIT OpenCourseWare) | .94 | 2.70 |
| 51 | Augmented reality application (e.g., Geotagging in Google Earth) | .42 | 2.63 |
| 13 | Learning software, educational software | .83 | 2.42 |
| 11 | Wikis with active participation as part of a course | .63 | 2.25 |
| 9 | E-learning applications as part of a course | 1.20 | 1.89 |
| 53 | Mobile apps for learning (e.g., iTunesU, iBooks) | .49 | 1.45 |

The instructor survey shows generally lower values of usage frequency in the e-learning application group; a minority stated using video sharing websites (on place 1 of this group of items, which paralleled students' ranking), e-learning applications as part of a course, and learning software / educational software (both of where were different than students' results), followed by recorded lectures (which was similar to students' ranking). Online tests and exams were in the middle field, but rarely utilized, and only very few instructors stated using mobile apps for learning, game-based learning applications, and augmented reality applications.

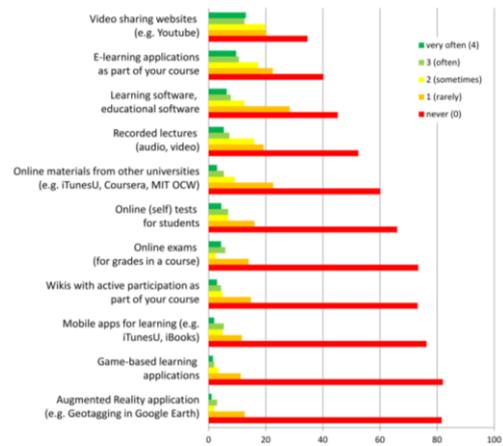

**Figure 10:** Valid percent of instructors' responses to the question: How often do you use the following for your academic work (i.e., teaching, research, service)?

**Table 9**. Instructors' responses to the question: How often do you use the following for your academic work (i.e., teaching, research, service)?

| | How often do you use the following for your academic work (i.e. teaching, research, service)? | Mean | Std. Dev. | valid n | (0) never | 1 (rarely) | 2 (some-times) | 3 (often) | (4) very often |
|---|---|---|---|---|---|---|---|---|---|
| 9 | E-learning applications as part of a course | 1.27 | 1.339 | 209 | 40.2 | 22.5 | 17.2 | 10.5 | 9.6 |
| 11 | Wikis with active participation as part of a course | .49 | .976 | 209 | 73.2 | 14.8 | 4.8 | 4.3 | 2.9 |
| 12 | Online materials from other universities (e.g., iTunesU, Coursera, MIT OpenCourseWare,) | .68 | 1.033 | 208 | 60.1 | 22.6 | 9.1 | 5.3 | 2.9 |
| 13 | Learning software, educational software | 1.01 | 1.206 | 208 | 45.2 | 28.4 | 12.5 | 7.7 | 6.3 |
| 23 | Recorded lectures (audio, video) | .94 | 1.204 | 208 | 52.4 | 19.2 | 15.9 | 7.2 | 5.3 |
| 36 | Online exams (for grades in a course) | .54 | 1.083 | 207 | 73.4 | 14.0 | 2.4 | 5.8 | 4.3 |
| 37 | Online (self) tests for studying | .67 | 1.137 | 206 | 66.0 | 16.0 | 6.8 | 6.8 | 4.4 |
| 47 | Video sharing websites (e.g., YouTube) | 1.49 | 1.407 | 208 | 34.6 | 20.2 | 19.7 | 12.5 | 13.0 |
| 50 | Game-based learning applications | .30 | .755 | 206 | 82.0 | 11.2 | 3.4 | 1.9 | 1.5 |
| 51 | Augmented reality application (e.g., Geotagging in Google Earth) | .29 | .734 | 206 | 81.6 | 12.6 | 1.9 | 2.9 | 1.0 |
| 53 | Mobile apps for learning (e.g., iTunesU, iBooks) | .45 | .948 | 207 | 76.3 | 11.6 | 4.8 | 5.3 | 1.9 |

## Frequency of Social Media Usage

Google search was the most commonly used web service by students for learning and study purposes, with Wikipedia as a moderately close second. Facebook was only in moderate use for learning, and Twitter and Google+ were quite infrequently used for this purpose.

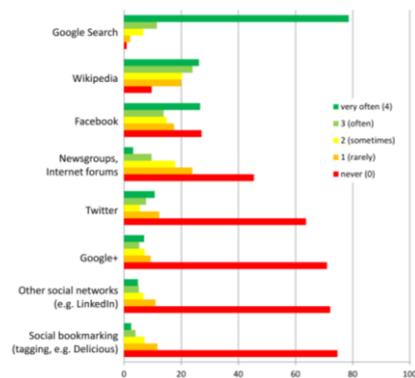

**Figure 11.** Valid percentage of students' responses to the question: How often do you use the following for learning/studying?

**Table 10.** Students' answers to the question: How often do you use the following for learning/studying?

| How often do you use the following for learning/studying? | | Mean | Std. Dev. | valid n | Valid Percent | | | | |
|---|---|---|---|---|---|---|---|---|---|
| | | | | | (0) never | 1 (rarely) | 2 (some-times) | 3 (often) | (4) very often |
| 10 | Newsgroups, Internet forums | 1.01 | 1.14 | 975 | 45.4 | 23.9 | 17.9 | 9.6 | 3.2 |
| 33 | Wikipedia | 2.37 | 1.32 | 961 | 9.7 | 19.9 | 20.3 | 23.9 | 26.2 |
| 40 | Google search | 3.65 | 0.78 | 962 | 0.9 | 2.2 | 6.8 | 11.5 | 78.6 |
| 42 | Social bookmarking and tagging ( e.g., Delicious) | 0.48 | 0.97 | 968 | 74.7 | 11.7 | 7.1 | 4.0 | 2.5 |
| 43 | Facebook | 1.95 | 1.57 | 966 | 27.1 | 17.5 | 14.9 | 13.9 | 26.6 |
| 44 | Google+ | 0.68 | 1.24 | 953 | 71.0 | 9.3 | 7.2 | 5.4 | 7.0 |
| 45 | Other social networks (e.g., LinkedIn) | 0.60 | 1.13 | 961 | 72.1 | 11.0 | 6.8 | 5.2 | 4.9 |
| 48 | Twitter | 0.90 | 1.4 | 958 | 63.7 | 12.3 | 5.5 | 7.7 | 10.8 |

## Satisfaction with the Usage of Social Media

The dominance of Google search is demonstrated not only in the values for the usage frequency, but also in the satisfaction results, which are slightly lower than the frequency but higher than all other social media variations. Twitter, social bookmarking, and other social networks have high satisfaction values despite low usage frequency. Facebook seems to be more frequently used, but not as satisfying as other social media at the moment of the survey. The relatively new Google+ application shows the lowest frequency and satisfaction values, but these values might increase in the next months as users become more acquainted with it.

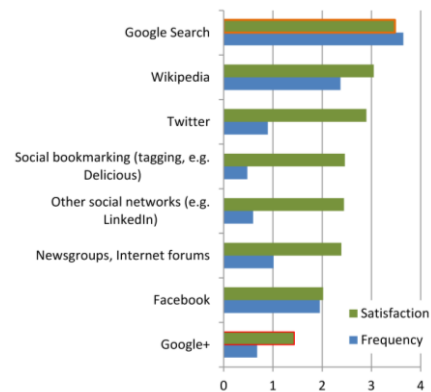

**Figure 12.** Means of students' responses to the questions: How often do you use the following for learning/studying? and If you use it: How satisfied are you with the use/functionality of the following for learning/studying? (red outline: in direction of dissatisfaction).

**Table 11.** Students' answers to the questions: How often do you use the following for learning/studying? and If you use it: How satisfied are you with the use/functionality of the following for learning/studying?

| How often do you use the following for learning/studying? and If you use it: How satisfied are you with the use/functionality of the following for learning/studying? | | Frequency | Satisfaction |
|---|---|---|---|
| 40 | Google search | 3.65 | 3.48 |
| 33 | Wikipedia | 2.37 | 3.05 |
| 48 | Twitter | .90 | 2.90 |
| 42 | Social bookmarking and tagging (e.g., Delicious) | .48 | 2.46 |
| 45 | Other social networks (e.g., LinkedIn) | .60 | 2.44 |
| 10 | Newsgroups, Internet forums | 1.01 | 2.39 |
| 43 | Facebook | 1.95 | 2.02 |
| 44 | Google+ | .68 | 1.43 |

The instructor survey shows a similar high usage frequency value for Google search and Wikipedia with a distance on place 2, which is the same as that in the student survey. Facebook had a lower rank, as did Twitter, as a result of which, Google+ increased by a few positions. Instructors might utilize cooperative applications in scientific projects with partners from different locations more frequently than students.

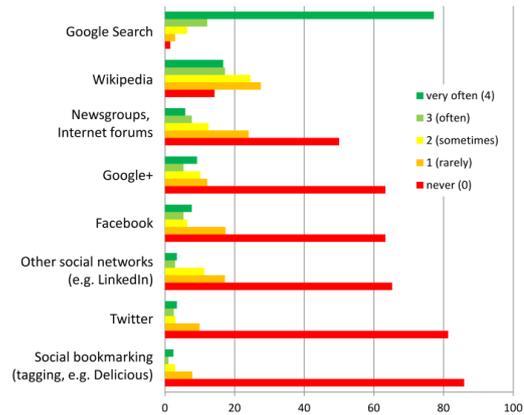

**Figure 13.** Valid percent of instructors' responses to the question: How often do you use the following for your academic work (i.e., teaching, research, service)?

**Table 12.** Instructors' responses to the question: How often do you use the following for your academic work (i.e. teaching, research, service)?

| | How often do you use the following for your academic work (i.e. teaching, research, service)? | Mean | Std. Dev. | valid n | Valid Percent | | | | |
|---|---|---|---|---|---|---|---|---|---|
| | | | | | (0) never | 1 (rarely) | 2 (some-times) | 3 (often) | (4) very often |
| 10 | Newsgroups, Internet forums | .95 | 1.207 | 208 | 50.0 | 24.0 | 12.5 | 7.7 | 5.8 |
| 33 | Wikipedia | 1.95 | 1.299 | 204 | 14.2 | 27.5 | 24.5 | 17.2 | 16.7 |
| 40 | Google search | 3.61 | .847 | 206 | 1.5 | 2.9 | 6.3 | 12.1 | 77.2 |
| 42 | Social bookmarking tagging (e.g., Delicious) | .26 | .777 | 206 | 85.9 | 7.8 | 2.9 | 1.0 | 2.4 |
| 43 | Facebook | .77 | 1.248 | 207 | 63.3 | 17.4 | 6.3 | 5.3 | 7.7 |
| 44 | Google+ | .85 | 1.326 | 207 | 63.3 | 12.1 | 10.1 | 5.3 | 9.2 |
| 45 | Other social networks (e.g., LinkedIn) | .62 | 1.027 | 204 | 65.2 | 17.2 | 11.3 | 2.9 | 3.4 |
| 48 | Twitter | .37 | .926 | 203 | 81.3 | 9.9 | 3.0 | 2.5 | 3.4 |

# 6. International comparison of media usage surveys

The survey at Western University followed the same concept as surveys in Europe and Asia. Nevertheless, an international comparison is problematic because the circumstances are very diverse and in dynamic change. In addition, the development could just be interpreted if repeated surveys have been conducted in different years. In this way, while it is risky to draw conclusions about international similarities and differences, some of the results may be correlated, such as comparing frequencies of e-learning application usage and comparing frequencies of social media usage.

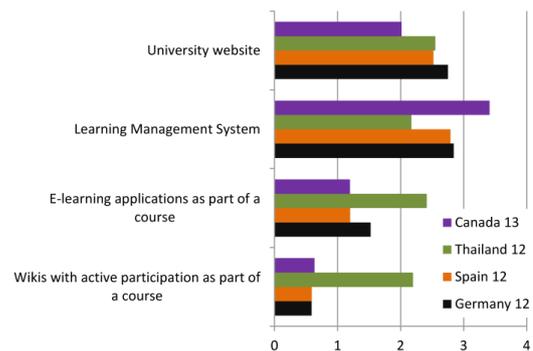

**Figure 14.** Students from four selected universities, one in Canada (valid n = 985), Germany (valid n = 1236), Spain (valid n = 981), and Thailand (valid n = 968), answered the question: How often do you use the following for learning/studying?. The question was rated on a five-point Likert scale with the following choices: never (0), rarely (1), sometimes (2), often (3), and very often (4) (or equivalent; the figure shows the means of all those who answered these questions).

**Table 13**. Students' answers to the question: How often do you use the following for learning/studying?

|  | Germany 12 | Spain 12 | Thailand 12 | Canada 13 |
|---|---|---|---|---|
| University website | 2.75 | 2.52 | 2.55 | 2.02 |
| Learning management system | 2.84 | 2.79 | 2.17 | 3.41 |
| E-learning applications as part of a course | 1.52 | 1.20 | 2.41 | 1.20 |
| Wikis with active participation as part of a course | .59 | .59 | 2.20 | .63 |

Looking at the means of four involved universities, it can be stated that the university websites were used slightly more frequently at the German university, followed by the Spanish university, the Thai university, and at a distance the Canadian university. The usage frequency of the learning management system was higher at this Canadian university; the Thai university showed the lowest frequency usage value in this item. There it seems to be more common – in comparison to these three other institutions – to utilize e-Learning applications as a part of a course and wikis with active participation as part of a course.

The utilization of Google search seems to be dominant in all the involved cases, and Wikipedia shows a certain relevance on a lower, but also remarkable, level in the survey results from all four universities (slightly higher at the German and the Thai universities). The results concerning the usage of video sharing websites like YouTube show a higher value in Thailand, followed by Spain and Canada; the lowest value is in the German case (in the item in the group of questions about "how often do you use the following for learning/studying," the results for free-time use are different).

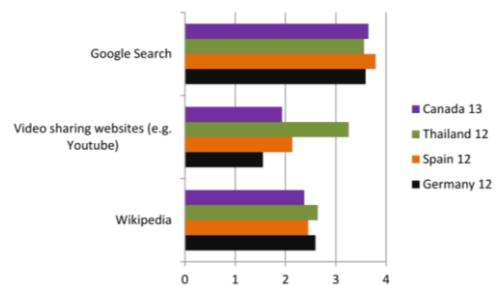

**Figure 15.** Students from four selected universities, one in Canada (valid n = 985), Germany (valid n = 1236), Spain (valid n = 981), and Thailand (valid n = 968), answered the question: How often do you use the following for learning/studying?)

**Table 14**. Students' answers to the question: How often do you use the following for learning/studying?

|  | Germany 12 | Spain 12 | Thailand 12 | Canada 13 |
|---|---|---|---|---|
| Google search | 3.59 | 3.79 | 3.56 | 3.65 |
| Video sharing websites (e.g., YouTube) | 1.55 | 2.14 | 3.25 | 1.93 |
| Wikipedia | 2.59 | 2.45 | 2.64 | 2.37 |

# 7. Discussion and implications

The results of this study support the assumption that the media usage of students and instructors includes a mixture of traditional and new media. The main traditional media continue to be important, and some new media have emerged on seemingly equal footing or are even more important than the traditional forms of media. Some new media that have recently been in the public spotlight do not seem to be as important as expected. These new media may still be emerging and it is not possible to know their ultimate importance at this point. There was some variation in media usage across different faculties, but perhaps not as much variation as might have been expected.

Of particular interest to one of primary co-investigator, a software engineer, was that engineering students were significantly different than their fellow students in their frequency of usage of a small number of media (e.g., computer games). Instructors showed some differences in their reported media usage, but there were notable similarities as well, such as the seemingly pervasive use of Google search.

Looking at the survey results, it can be stated that several **traditional media** were still very relevant and continued to be in high use, even in the context of a changing environment. Printed material and slides from instructors as well as printed books were deemed to have high values of usage frequency and satisfaction. Attending class and visiting libraries were frequently performed habits, and the universities' services were used more frequently than external academic sources.

At the same time, **additional new media**, such as electronic versions of material from instructors or the learning management system, were established and utilized

**EAI** European Alliance for Innovation

with a similar intensity. It seems that these newly established media, which are based on traditional media, are very easy and comfortable to access and use and, therefore, in the future they are likely to be used more often than their traditional counterparts.

This intensive use of new media services and arrangements might be a phenomenon enabled by **new habits** that encourage working with media. Students and instructors are equipped with mobile and continuously network-connected computers, and they are proficient in using them based on the experience (and self-organized learning) in their private life. The use of some media can be understood as obligatory, especially the use of Google search, which had the highest rank in both usage frequency and satisfaction values. Differences in usage exist between students and instructors and between free time and studying usage. The use of Facebook and YouTube show very high values of usage frequency, so might also be classified as habits.

Certain innovative usage **variations of new media** for teaching and learning/studying are distinct, such as wikis as a part of a course, recorded lectures, or online tests, but more often for certain courses. Wikis have been developed and launched, and their effectiveness has been proven; however, just a few arrangements seem to apply to these options. It can be assumed that in the cases where a serious effort has been made, these new variations of working with new media have a distinct relevance, such as recorded lectures in science courses.

Media usage expands the interdependence with the market of academic education. As a result, **competition** with other universities and service providers has intensified. Although the frequency of use of online materials from other universities (e.g., iTunesU, Coursera, MIT OpenCourseWare) or mobile apps for learning has not reached a similar level as the use of materials from Western University, the use of media with a non-direct competitive influence seems to be especially remarkable, such as video sharing websites, Wikipedia, or Google Books. It can be assumed that the competition will be much more intense in the future, as the main players on the market continue to collect (and utilize) much more specific data about students and instructors than any single university are able, or would be permitted, to do.

Potentially arising **future media and trends** cannot be identified with this survey, but relatively new media, such as Google+, augmented reality applications, or game-based learning applications, might become more important for teaching and studying, although they are not currently in common use. In addition, the side effects of some of the established and ubiquitous usage of some media will likely have consequences. In this way, working with Google search facilitates so-called "hyper targeting" and creating electronic user profiles that can be used for technology-based customization and the delivery of services at a high level of situational individualization.

Overall, the media usage by **students and instructors**, while different in some aspects, is explainable, as in the case of desktop PCs, Facebook, and YouTube. Instructors, as a heterogeneous group, generally had a more traditionally oriented usage of media, but some showed ingenuity in using new options. In this way, the frequency of using Google+ was higher for instructors than students. Many new media were used extensively by both instructors and students and can be considered as "new habits" (in a world of academia, where some habits seem to be unchangeable, although that has been intended over the years).

The students' age and years of **study experience** doesn't appear to make significant differences in the media usage frequency between freshmen and senior students. This particular component is complex because students are a heterogeneous group. Some study habits may be a result of the duration of study, but others habits that are walking in the opposite direction, may have been influenced by different experiences from their adolescence.

Students from different **faculties** show a general similarity. Significant differences can be noted in the comparison between two different faculties, such as arts and humanities vs. science or engineering (e.g., with the frequency of reading books in the arts and humanities), but this is explainable. Additionally, gender has a significant influence, especially in the frequency of use of "social media."

The survey at Western University followed the same concept as surveys conducted in Europe and Asia. An **international** comparison is problematic; the development could just be interpreted if repeated surveys have been conducted. In this way, it is speculative to answer questions about international similarities and differences. Nevertheless, it seems that the usage of IT devices might differ (e.g., more smartphone usage in Thailand/Asia and even Germany/Europe compared to Canada) and the use of social media in academic education seems to be more common in Thailand/Asia compared to Canada and Germany. The competitiveness of the Internet-based market of academic education might be more intensive in Canada because of the proximity of the U.S. market.

## Implications of the Results for Main Teaching Formats

It might be relevant to think about optional consequences of the survey results for the media usage in the main teaching formats at the university – lecture-based courses and seminar/project-based courses – as well as infrastructural arrangements the university prepares and offers to students and instructors.

## Lecture-Based Courses

Lecture-based courses are still the most relevant and common teaching format at the university. The advantage of direct interaction and guidance seems to be continuously appreciated and to work successfully with established (traditional) media. In addition, several new media are in use, such as the electronic counterparts of the printed media and applications that facilitate course routines (e.g., learning management systems).

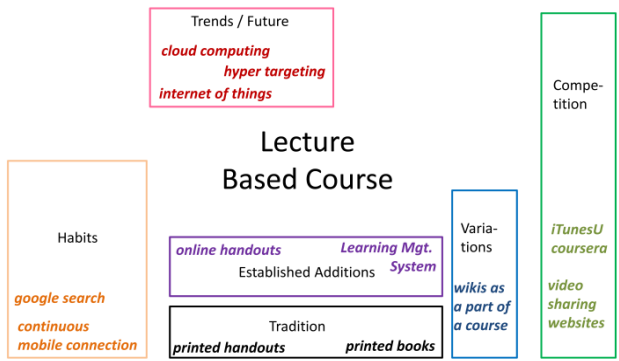

**Figure 16.** Usage of media in relation to a lecture-based course

Lecture-based courses could (and should) utilize students' new habits of studying with media. In this way, it can be assumed that students are equipped, connected, positive towards, and basically proficient in working with new media. It seems to be quite possible to enhance lecture-based courses with various arrangements that involve new media. In the field of this teaching format, the competition at the market seems to be most active, when prepared and video sharing websites offer attractive presentations and explanations of difficult and abstract contents become an available alternative to the lecture of every instructor. The local relation of students and instructors – if positively developed – will perhaps be influenced by the ubiquitous presence of other actors that are connected through media.

## Seminar- and Project-Based Courses

Seminar- and project-based courses are usually characterized by student activities and common work with new and open tasks, didactic use of group initiative, and student knowledge. Students seem to be very open and prepared to use tools that support these elements. In addition, media tools could offer methods to supervise and coach student teams. The heterogeneity of both students and instructors might lead to decisions to not use just a few media arrangements, but to flexibly use diverse variations – like the range of media that can be seen in the survey. Regardless of whether or not these media are utilized every instructor will have the opportunity to use additional media applications that address students' needs directly or are brought by student into the classroom. In principle, these media add the option of combining classroom arrangements with other learning locations and enable the process of developing an international academic education.

## Infrastructure

The dynamics surrounding the field of media and the often unconditional willingness to perform risky activities on the Internet pressure the university to connect its own systems to the outside network so that it can offer comfortable and modern (innovative) services and applications. Nevertheless, students (and instructors) still seem to esteem and expect the university to be a solid and responsible actor. Identification with a certain university and its campus (life) is important and can be the reason that a university's own media are used more frequently than external options. A combination of solid, secure, controllable internal media infrastructures, active connections and cooperation with partners and providers on the market, and innovative variations on trial, along with the support of the university's competent media experts, would help universities to find a unique position in the competitive market of academic education and scientific research.

## 8. Conclusions and recommendations

The survey results might be considered in four domains of the media activity at universities: (1) investment and development of basic arrangements, services, and infrastructure; (2) strategic planning concerning media usage for teaching and studying; (3) support for innovative projects to test and establish new media applications in academic education; and (4) active utilization of external arrangements and services through cooperation and utilization.

Concerning investment and development of **basic arrangements**, services, and infrastructure, the survey results show the importance of each university's individual competence, system, and responsibility (in comparison with external options and services). In cases where the university decided to install and establish certain media, such as the learning management system, computer laboratories, or recorded science lectures, this media had a distinct relevance to this field.

Concerning **strategic planning** to develop media usage for teaching and studying, the survey gives information about the actual situation and ideas about future trends. It seems to be obvious that ubiquitous habits should be recognized and their consequences and effects be considered, as in the case of Google search (e.g., the company might change, but the phenomenon will continue). Although media may not be a central factor of universities' strategies, the decisions concerning the support of certain media should be related to main strategies (e.g., internationalization).

Concerning support for **innovative projects** to test and establish new media applications in academic education, the survey shows the spectrum of currently relevant media and shows that their success depends to a certain extent on official and serious support. The dynamic change of media usage habits requires the active testing and fostering of usage of new media for teaching and studying (such as wikis as a part of a course, Google+, or mobile apps for learning) and the continuous support for activities to join main media arrangements in academic education on the open market (such as video sharing websites and virtual classrooms).

In regards to active utilization of external arrangements and services through **cooperation**, it seems to be unavoidable to take the usage habits of students (and instructors) as a fact and integrate available services and offerings into teaching (such as materials from other universities). A recommendation would be to join some of the already existing communities or to initiate or reinforce organized cooperation with other specific universities.

## References

[1] L. Johnson, S. A. Becker, V. Estrada & A. Freeman, NMC Horizon Report: 2014 Higher Education Edition, New Media Consortium: Austin, Texas, 2014.

[2] E. Dahlstrom, ECAR Study of Undergraduate Students and Information Technology, EDUCAUSE Center for Applied Research: Louisville, CO, USA, http://net.educause.edu/ir/library/pdf/ERS1208/ERS1208.pdf, 2012.

[3] D. Buckingham, Media Education Goes Digital: An Introduction, Journal of Learning, Media and Technology, 32(2):111-119, Routledge, DOI: 10.1080/17439880701343006, June 2007.

[4] I. Marenzi & S. Zerr, Multiliteracies and Active Learning in CLIL - The Development of LearnWeb2.0, IEEE Transactions on Learning Technologies, 5(4):336-348, DOI: 10.1109/TLT.2012.14, 2012.

[5] M.A. Chatti, U. Schroeder & M. Jarke, LaaN: Convergence of Knowledge Management and Technology-Enhanced Learning, IEEE Transactions on Learning Technologies, 5(2):177-189, DOI: 10.1109/TLT.2012.33, 2012.

[6] M. Alrasheedi & L.F. Capretz, A Meta-Analysis of Critical Success Factors Affecting Mobile Learning, Proc. of IEEE International Conference on Teaching, Assessment and Learning for Engineering (TALE), Bali, Indonesia, pp. 262-267, IEEE Press, August 2013.

[7] S.D. Smith, G. Salaway & J.G. Caruso, The ECAR Study of Undergraduate Students and Information Technology, EDUCAUSE, 2009

[8] C.C. Pritchett, E.C. Wohleb & C.G. Pritchett,Educators' Perceived Importance of Web 2.0 Technology Applications. TechTrends, 57 (2):33-38, 2013.

[9] A. Murphy, H. Farley, M. Lane, A. Hafess-Baig & B. Carter, Mobile Learning Anytime, Anywhere: What Are our Students Doing?, 24th Australasian Conference on Information Systems, 11 pages, 2013.

[10] L. Lockyer & J. Patterson, Integrating Social Networking Technologies in Education: a Case Study of a Formal Learning Environment, 8th IEEE International Conference on Advanced Learning Technologies, pp. 529-533, DOI: 10.1109/ICALT.2008.67, 2008.

[11] G.M. Johnson & J.A. Johnson, J.A. Dimensions of Online Behavior: Implications for Engineering E-Learning, in Technological Developments in Education and Automation (M. Iskander et al., eds), 6 pages, DOI: 10.1007/978-90-481-3656-8-13, 2010.

[12] M. Cusumano, Are the Costs of 'Free' Too High in Online Education? Communications of the ACM, 56(4):1-4, DOI: 10.1145/2436256.2436264, 2013.

[13] T.R. Klassen, Upgraded Anxiety and the Aging Expert, Academic Matters – Journal of Higher Education, Ontario Confederation of University Faculty Association, pp. 7-14, May-June 2012.

[14] A.S. Kazley, D.L. Annan, N.E. Carson, M. Freeland, A.B. Hodge, G.A. Seif & G.S. Zoller, Understanding the use of Educational Technology Among Faculty, Staff, and Students at a Medical University. TechTrends, 57 (2):63-70, 2013.

[15] G. Gidion, L.F. Capretz, G. Grosch & K. Meadows, Are students satisfied with media:a Canadian cases study, Bulletin of the IEEE Technical Committee on Learning Technology, 16(1):6-9, January 2014.

[16] G. Gidion, L.F. Capretz, G. Grosch & K. Meadows, Media Usage Survey: How Engineering Instructors and Students Use Media, Proceedings of the Canadian Engineering Education Association Conference (CEEA'13), pp. 1-5, July 2013.

[17] M. Alrasheedi & L.F. Capretz, An M-Learning Maturity Model for the Educational Sector, 6th Conference of MIT Learning International Networks Consortium (MIT LINC), Boston, MA, pp. 1-10, June 2013.

[18] A. Ali, A. Ouda & L.F. Capretz, A Conceptual Framework for Measuring the Quality Aspects of Mobile Learning, Bulletin of the IEEE Technical Committee on Learning Technology, 14(4):31-34, IEEE, October 2012.

[19] M. Grosch & G Gidion, Mediennutzungsgewohnheiten im Wandel (German). Ergebnisse einer Befragung zur studiumsbezogenen Mediennutzung, KIT Scientific Publishing, http://digbib.ubka.uni-karlsruhe.de/volltexte/1000022524, 2011.

[20] M. Grosch, Mediennutzung im Studium. Eine empirische Untersuchung am Karlsruher Institut für Technologie, Shaker, 2012.

[21] L.J. Sax, S.K. Gilmartin & A.N. Bryant, Assessing Response Rates and Non-response Bias in Web and Paper Surveys, Research in higher education,44(4): 409-432, 2009.